\documentclass[epj]{svjour}
\usepackage{epsfig}
\usepackage{graphicx}
\usepackage{amssymb, amsmath}
\usepackage{version}
\usepackage{float}
\usepackage{color}
\usepackage{array}
\newcommand{\ddr}{\textrm{d}}
\newcommand{\dd}{\textrm{d}}
\newcommand{\im}{\textrm{i}}
\newcommand{\vvec}{{\bf v}}
\newcommand{\cvec}{{\bf c}}
\newcommand{\fvec}{{\bf F}}
\newcommand{\rvec}{{\bf r}}
\newcommand{\Fvec}{{\bf F}}
\newcommand{\bv}{{\bf v}}

\newcommand{\bF}{{\bf F}}

\newcommand{\svec}{\hat{\boldsymbol \sigma}}
\graphicspath{{figures/}}

\begin{document} 
\onecolumn
\title{Energy fluctuations in vibrated and driven granular gases}
\date{\today} 

\author{Paolo Visco \inst{1,2} \and Andrea Puglisi \inst{1} \and Alain Barrat
  \inst{1} \and Fr\'ed\'eric van Wijland \inst{1,3} \and Emmanuel Trizac
  \inst{2}}

\institute{Laboratoire de Physique Th\'eorique (CNRS
  UMR8627), B\^atiment 210, Universit\'e Paris-Sud, 91405 Orsay cedex, France
  \and 
  Laboratoire de Physique Th\'eorique et Mod\`eles
  Statistiques (CNRS UMR8626), B\^atiment 100, Universit\'e Paris-Sud, 91405
  Orsay cedex, France
  \and
  Laboratoire 
  Mati\`ere et Syst\`emes Complexes (CNRS UMR 7057), Universit\'e
  Denis Diderot (Paris VII), 2 place Jussieu, 75251 Paris cedex 05,
  France}

\abstract{
  We investigate the behavior of energy fluctuations in several models of
  granular gases maintained in a non-equilibrium steady state.  In the case of
  a gas heated from a boundary, the inhomogeneities of the system 
  play a predominant role. Interpreting the total kinetic energy as a
  sum of independent but not identically distributed random variables, it is
  possible to compute the probability density function (pdf) of the total
  energy. Neglecting correlations and using 
  the analytical expression for the inhomogeneous temperature profile obtained
  from the granular hydrodynamic equations, we recover results which have been
  previously observed numerically 
  and which had been attributed to the presence of
  correlations. In order to separate the effects of spatial inhomogeneities
  from those ascribable to
  velocity correlations, we have also considered two models of
  homogeneously thermostated gases: in this framework it is possible to 
  reveal the presence of non-trivial effects due to velocity
  correlations between particles. Such correlations stem from the
  inelasticity of collisions. Moreover, the observation that the pdf of the
  total energy tends to a Gaussian in the large system limit, suggests that
  they are also due to the finite size of the system.}

\maketitle
\section{Introduction}
In non-equilibrium statistical mechanics the emergence of non-Gaussian
distributions is a remarkable feature that marks an
essential difference with typical results of equilibrium statistical
mechanics. In particular, non-Gaussian behaviors for {\em global} quantities
(i.e. system averaged) have been observed in different fields of
physics \cite{racz03}, unveiling unexpected analogies between turbulent flows,
equilibrium critical phenomena \cite{bramwell98}, and non-equilibrium
instabilities \cite{brey05}.  Averaged quantities are expected to be free of
the microscopic details of the system under study, in such a way that their
probability distribution function (pdf) should only depend on few
parameters. Furthermore, they will contain relevant informations about the
correlations among different parts of the system.

Granular gases represent a simple example of many-particles system out
of equilibrium. The simplest model consists in an assembly
of smooth hard particles undergoing binary inelastic collisions. 
This model, known as the Inelastic Hard Spheres (IHS)
model, has stimulated a great interest for more than ten years.
Its close similarity with the elastic hard spheres fluid has made
possible the exploitation of the long-past known standard methods of
kinetic theory. However the inclusion of energy dissipation in the
collision law is sufficient to generate many surprising features that
have been observed both in experiments and simulations, such as
clustering, convection, non-equipartition of energy, {\it etc.}
\cite{barrat05,poschel01,poschel03}. 

In this paper we study the pdfs of the total kinetic energy of granular gases
maintained in a non-equilibrium steady state by an external driving mechanism
which compensates for the energy loss due to collisions.  We first focus on
boundary driven gases, since this kind of driving mechanism can easily be
achieved experimentally, vibrating at high frequencies the container of the
gas or one of its walls \cite{losert99b,rouyer00,kudrolli00,wildman02}.
Energy is hence provided at the boundaries and dissipated by inelastic
collisions in the bulk of the system, leading to strong spatial
inhomogeneities. A coarse-grained description of this steady state, in terms
of density and temperature profiles, can be obtained by means of hydrodynamic
equations~\cite{brey00c}, which are {\it a priori} valid when the
mean-free-path is small compared to the typical length scale of the gradients.
The presence of non-uniform density and temperature profiles implies that the
total energy is a sum of variables which are not identically distributed and,
in principle, correlated. We will first investigate the consequences of the
lack of homogeneity by assuming total independence among the velocities, i.e.
absence of correlations. Interestingly enough, this approximate approach leads
to good estimates for the shape of the pdfs. Similarly, Bertin in a recent
paper~\cite{bertin05} showed that considering the sum of independent, but not
identically distributed, random variables can be a useful way to explain (at
least on a qualitative level) the emergence of non-Gaussian pdfs in other
systems.

We will also discuss another important characteristic of such
inhomogeneous systems, i.e. the fact that they are often
non-extensive. Since the energy is injected from the boundaries and
the dissipation happens inside the volume, it is clear that increasing
the volume at constant density will not {\it a priori} increase the
total energy linearly with the total number of particles $N$. The
hydrodynamic approach allows to show that the relevant variable 
is the number of layers of particles at rest \cite{brey00c}.

In order to disentangle the effects of inhomogeneities from those of
correlations, we focus, in the second part of the paper, on the energy
fluctuations in homogeneous granular gases. We will enforce homogeneity by
considering the system in a pure velocity space, i.e. eliminating any spatial
degree of freedom. When inhomogeneities disappear, the effects that may
contribute, in the thermodynamic limit, to a non-Gaussian pdf of the total
energy are the correlations among velocities. Nonetheless such a behavior
generally shows up when the system is near some critical point, i.e. when some
correlation length (in velocity space) diverges. A general theory for
correlation functions for the hard sphere fluid has been first derived in
\cite{ernst81}, and fruitfully applied to the homogeneous cooling state (HCS)
for a gas of inelastic hard spheres~\cite{brey04a}. This last result was
obtained exploiting the fact that the dynamics is dominated by the
hydrodynamic (i.e. slow) modes, which are exactly computable for the
particular case of the HCS. In our homogeneous models the total (kinetic)
energy is the sum of $N$ identically distributed random variables. We will
show, by numerical simulations, that in this state the central limit theorem
applies and the energy pdf is a Gaussian in the large $N$ limit, which means
that the correlation length in the velocity space is finite.  The interesting
issue then concerns the computation of the variance of the rescaled pdf. This
quantity depends on the one-particle and (with a pre-factor $1/N$)
two-particles distribution functions. It is deeply related to the finite size
correlations between the velocities of the particles. In both cases of a
stochastic and of a deterministic thermostat, its measure reveals the presence
of non-trivial correlations. Remarkably, it only depends upon the restitution
coefficient, but not upon other parameters such as the number of particles or
the amplitude of the thermostating force. In the deterministic thermostated
case, it is moreover in almost perfect agreement with the theoretical
computations given in~\cite{brey04a} for a gas of inelastic hard spheres.

The structure of the paper is the following: in section II we 
discuss the gas driven at the boundaries, while in section III we 
consider the homogeneous thermostats. Conclusions are drawn in
section IV.

\section{The inhomogeneous boundary driven gas}
The focus is here on the energy fluctuations of a granular gas in the
case where energy is injected through a boundary, typically a
vibrating wall. This kind of system corresponds to a realistic
experimental setup \cite{losert99b,rouyer00,kudrolli00,wildman02} and
has therefore been widely studied both numerically and analytically
\cite{grossman97,mcnamara97,mcnamara98b,kumaran98,brey00c,barrat02b}.
As an important consequence of the boundary heating, the density and
the temperature are not homogeneous: a heat flux is present, which
does not verify the Fourier law. This feature is well described by
kinetic theory and in good agreement with the hydrodynamic
approximation, which allows an analytical calculation of the density
and temperature profiles~\cite{brey00c}.  Recently Auma\^itre {\it et
  al.}  \cite{aumaitre04} investigated, by means of Molecular Dynamics
(MD), the fluctuations of the total energy of the system. In
particular they have studied the behavior of the first two moments of
the energy probability distribution function (pdf) when the system
size is changed, at constant averaged density. Because of the
inhomogeneities, the mean kinetic energy is no more proportional to
the number of particles, and thus it is not an extensive quantity;
analogously the mean kinetic temperature, defined as the average
kinetic energy per particle, is no more intensive. This has
led~\cite{aumaitre04} to the proposal of a definition of an effective
temperature $T_E$ and an effective number of particles $N_f$, such
that $T_E$ is intensive and the total mean energy $\langle
E\rangle=N_f T_E$ is extensive {\em in this effective number of
  particles}.  In the following we will show how an approximate
calculation {\em which neglects correlations and the small
  non-Gaussianities}, using the hydrodynamic prediction of
\cite{brey00c} for the temperature profile, allows in fact to explain
the phenomenology observed in \cite{aumaitre04}. Within this
description it is possible to obtain an explicit expression of the
effective temperature and effective number of particles as a function
of the system parameters (i.e. number of particles, restitution
coefficient, and temperature of the vibrating wall). At large enough
system sizes, the effective temperature is shown to be independent of
the system size, while the effective number of particles is
proportional to the surface of the heating boundary.

\subsection{Energy Probability Distribution Function}

We consider a granular gas of $N$ smooth inelastic hard spheres of
diameter $\sigma$ between two parallel walls. The distance between the
two walls is denoted by $H$, oriented along the $x$ axis.  The linear
extension of the wall is denoted by $L$, and periodic boundary conditions
are applied along the directions parallel to the wall. One of the
walls (in $x=0$) is vibrated, in order to compensate for the energy lost
through inelastic collisions. As usual in the IHS model,
the total momentum is conserved in
collisions, and only the normal component of the velocity is
affected. Thus, the collision law for a pair of particles $(1,2)$
reads:
\begin{equation}
\label{coll}
\begin{cases}
  {\bf v}_1^*={\bf v}_1-\frac{1}{2} (1+\alpha)({\bf v}_{12} \cdot {\hat
    {\boldsymbol \sigma}})
  {\hat {\boldsymbol \sigma}} \\
  {\bf v}_2^*={\bf v}_2+ \frac{1}{2} (1+\alpha)({\bf v}_{12} \cdot {\hat
    {\boldsymbol \sigma}})
  {\hat {\boldsymbol \sigma}}  \,\, ,\\
 \end{cases}
 \end{equation}
 where $\svec$ is a unitary vector along the center of the colliding particles
 at contact. In the dilute limit, such a system is well described by the
 Boltzmann equation, involving the one-particle distribution function $f(\rvec,
 \vvec, t)$:
\begin{equation}
\label{boltzpos}
\partial_t f(\rvec,\vvec_1, t)+ \vvec \cdot {\boldsymbol \nabla} 
f(\rvec,\vvec_1, t) = J[f|f]\,\,.
\end{equation}   
Here $J[f|f]$ is the collision integral, which takes into account the
inelasticity of the particles:
\begin{equation}
\label{collint}
J[f|f]=\sigma^{d-1} \int \ddr \vvec_2 \int_{\vvec_{12} \cdot \svec >
  0} \ddr \svec (\vvec_{12} \cdot \svec)
\left(\frac{f(\rvec,\vvec_1^{**},t)f(\rvec,\vvec_2^{**},t)}{\alpha^2}
  - f(\rvec,\vvec_1,t) f(\rvec,\vvec_2,t) \right) \,\,,
\end{equation}
where a collision transforms the couple ($\vvec^{**}_i,\vvec_j^{**})$
into $(\vvec_i,\vvec_j)$. The hydrodynamic fields (local density $n$,
local average velocity field ${\bf u}$, local temperature $T$) are
defined as the velocity moments:
\begin{subequations}
\label{hydrofields}
\begin{align}
n(\rvec,t) &=\int \ddr \vvec f(\rvec,\vvec,t)\,,\\
n(\rvec,t) {\bf u}(\rvec,t) & = \int \ddr \vvec \, \vvec f(\rvec,\vvec,t) \,,\\
\frac{d}{2} n(\rvec,t) T(\rvec,t) &= \int \ddr \vvec \frac{m}{2} 
(\vvec - {\bf u})^2 f(\rvec,\vvec,t)\,,
\end{align}
\end{subequations}
where $d$ is the space dimension. The hydrodynamic balance equations for the
afore-defined fields are derived taking the velocity moments in
(\ref{boltzpos}).  Then, it is possible to show that a steady state solution
without macroscopic velocity flow exists: this solution is invariant for
translations along all directions perpendicular to $x$, so that the
hydrodynamic fields only depend  upon $x$. Moreover this solution is stable if
the size (in the directions perpendicular to $x$) of the system is smaller
than a critical size which depends upon $\alpha$. In the rest of this section
we consider such a ``horizontally homogeneous'' solution, whose temperature
profile reads \cite{brey00c}:
\begin{equation}
\label{tempprofile}
T(\ell)=T_0 \left( \frac{\cosh {\left( \sqrt{a(\alpha)} (\ell_m- \ell) 
\right)}} {\cosh{\left( \sqrt{a(\alpha)} \ell_m \right)}} \right)^2 \,\,,
\end{equation}
where $a(\alpha)$ is a function of the restitution coefficient (its complete
expression is given in ref. \cite{brey00c}). The variable $\ell$ is
proportional to the integrated density of the system over the $x$ axis. Its
definition is given by the following relation involving the local
mean-free-path $\lambda (x)$:
\begin{equation}
\label{lscale}
\textrm{d} \ell=\frac{\textrm{d} x}{\lambda (x)} \,\,,\;\;\;\;\;\; 
\lambda(x)= \left[ \frac{\sqrt{2} \pi^{\frac{d-1}{2}}}{{\Gamma[(d+1)/2]}} 
\sigma^{d-1} n(x) \right]^{-1}\,\,,
\end{equation}
with $\ell=0$ at $x=0$.
On the other hand, $\ell_m$ is the maximal $\ell$-value,
reached at $x=H$:
\begin{equation}
\ell_m = C \sigma^{d-1} \, \int_0^H \textrm{d}x \,\, n(x) = C \sigma^{d-1} N_x
\,\,, 
\end{equation}
where $N_x$ is the number of particles per unit section
perpendicular to the $x$ axis (e.g. $N_x = N/L$ for $d=2$), and $C$ is
a constant depending on the dimension (e.g. $C=2 \sqrt{2}$ for $d=2$).
It thus turns out that the crucial quantity $\ell_m$ which determines
the hydrodynamic profiles is 
proportional to the {\em number of layers} of particles at rest.
The boundary conditions used to get expression (\ref{tempprofile}) are:
\begin{equation}
T(\ell=0)=T_0 \,\,,\,\,\,\,\,\,\,\,\,\,
\left.\frac{\partial T}{\partial \ell}\right|_{\ell=\ell_m}=\,\,
\left.\frac{\partial T}{\partial x}\right|_{x=H}=0\,\,,
\end{equation}
where we have assumed that the vibrating wall acts as a thermostat
that fixes to $T_0$ the temperature at $x=0$. A detailed observation
would show that this is not a very realistic assumption.
Moreover, the region in the vicinity of $x \sim 0$ is the one in which
the hydrodynamic description may break down. Nevertheless, $T_0$ may
be seen as an effective parameter such that
(\ref{tempprofile}) coincides with the observed profile in the region of
the system where hydrodynamics holds. In the following we will suppose
the velocity distribution to be a Maxwellian at each fixed distance
from the wall (non-Gaussian features may be observed, but are 
not relevant for the present calculation), with a
local temperature (variance) given by (\ref{tempprofile}):
\begin{equation}
f(\bv,\ell)={e^{-{v^2 \over 2 T(\ell)}} \over \sqrt{2 \pi T(\ell)}} \,\,.
\end{equation}
The distribution for the energy of one particle ($e=v^2/2$) is hence:
\begin{equation}
p(e,\ell)=f_{{1 \over T(\ell)}, {d \over 2}}(e) \,\,\,,
\end{equation}
where $f_{\alpha, \nu} (x)$ is the gamma distribution \cite{feller}:
\begin{equation}
\label{gamma}
f_{\alpha, \nu}(x)={\alpha^{\nu} \over \Gamma(\nu)} 
x^{\nu-1} e^{- \alpha x}\,. 
\end{equation}
The characteristic function of the gamma distribution is:
\begin{equation}
{\tilde f}_{\alpha,\nu}(k)= \int_{-\infty}^{+ \infty} \dd x e^{\im k x} 
f_{\alpha, \nu}(x) = \left(1-\frac{\im k}{\alpha} \right)^{- \nu}\,\,.
\end{equation}
Our interest goes to the macroscopic fluctuations integrated over all the
system. Thus, the macroscopic variable of interest is the granular temperature
$T_g$, defined here as the average of the local temperature over the $x$
profile:
\begin{equation}
\label{tgran}
  T_g={1 \over N}\int_V  n(\rvec) T(\rvec) \,\,\dd \rvec=
{1 \over \ell_m} \int_{0}^{\ell_m} T(\ell) \,\,\textrm{d} \ell \,\,.
\end{equation}
To obtain an expression of the energy pdf over all the system, it is useful
to divide the box in $\ell_m / \Delta \ell$ boxes of equal height (in
the $\ell$ scale) $\Delta \ell$. The use of this length scale allows
to have a fixed number of particle $N_{\ell}$ in each box of size
$L \times \Delta \ell$, since $d\ell \propto n(x)dx$. 
Moreover, in each box $i$ 
the temperature can be taken as uniform, denoted
$T_i \equiv T(i \,\, \Delta \ell)$.
In the $j-th$ box, we can calculate the energy $\epsilon_j$ which is
the sum of the energies of the $N_\ell$ particles in the box. The pdf
of this energy, $q_j(x)=\textrm{prob}(\epsilon_j=x)$, can be
straightforwardly computed when the velocities of the particles are
supposed to be uncorrelated:
\begin{equation}
q_j(x)=f_{{1 \over T_j}, N_{\ell} \frac{d}{2}}(x) \,\,\,,
\end{equation}
Thus, the characteristic function for the total kinetic energy $E= \sum
\epsilon_j$ can be obtained as the product of the characteristic function
${\tilde q}_j(k)$ of the pdf of each $\epsilon_j$:
\begin{equation}
\label{prod}
{\widetilde P}(k)= \prod_{j=0}^{\ell_m / \Delta \ell} {\tilde q}_j(k)=
\prod_{j=0}^{\ell_m / \Delta \ell} (1 - \im k
  T_j)^{- \frac{N_{\ell} d}{2}} \,\,\,.
\end{equation}
Since the number of particle per box $N_{\ell}$ is a known fraction of the
total number of particles ($N_{\ell} = N \Delta \ell / \ell_m$), one can
rewrite the expression (\ref{prod}) as a Riemann sum. In the limit $\Delta
\ell \to 0$ this yields the total kinetic energy characteristic function:
\begin{equation}
\label{charfunen}
{\widetilde P}(k)= \exp  \left(\, -{ N d\over 2 \ell_m } \int_0^{\ell_m} \log
\left(1-\im k T(\ell)\right) \,\,  \textrm{d} \ell \right)\,\,.  
\end{equation}
Note that this result is valid for any temperature profile $T(\ell)$ and hence
it can be applied also to other situations with different boundary conditions
or different hydrodynamic equations.

\subsection{Local energy fluctuations}

While we have focused on the pdf of the global energy in the previous
subsection, it is also interesting to consider an intermediate scale
and to compute the energy pdf for a "slice" of the system at a given
distance $x$ from the vibrating wall, and of small height $\Delta x$
(such that the local density and temperature can be considered as
uniform).  We still consider that the velocities are Gaussian and
uncorrelated. In this case the energy $\epsilon$ is a sum of $N(x)$
identically distributed random variables, but their number $N(x)$
fluctuates in time. Hence the pdf of the local energy is:
\begin{equation}
\label{eq:pdfstripe1}
P_{l}(\epsilon,x)=\sum_{N(x)=1}^{\infty} P_{d}(N(x)) f_{\frac{1}{T(x)},N(x)
  \frac{d}{2}} (\epsilon) +   P_d(0) \delta(\epsilon)\,\,, 
\end{equation}
where $P_{d}$ is the distribution of the number of particles at a
distance $x$ from the heated wall. The second term in the above
expression is due to the fact that when there are no particles the
energy pdf is a Dirac distribution. If the particles are uncorrelated
the distribution $P_{d}$ should be well approximated by a Poisson
distribution with average $\langle N(x) \rangle = n(x)\Delta x$:
\begin{equation}
P_{d}(N(x)=i)=e^{-n(x)\Delta x} \frac{(n(x)\Delta x)^i}{i !}\,\,.
\end{equation}
In this framework the explicit expression for the probability of the
local energy is given, in two dimensions, by:
\begin{equation}
\label{eq:pdfstripe2}
P_{l}(\epsilon,x)=e^{-n(x)\Delta x - \frac{\epsilon}{T(x)}} \
\sqrt{\frac{n(x)\Delta x}{T(x) \epsilon}} \,  I_1
\left(2 \sqrt{\frac{n(x)\Delta x\ \epsilon}{T(x)}} \right) 
+ 2 e^{-n(x)\Delta x} \delta(\epsilon)\,\,,
\end{equation}
where $I_n(x)$ is the first kind modified Bessel function
\cite{abramowitz}.  For higher dimensions the sum in
Eq.~(\ref{eq:pdfstripe1}) with Poissonian density fluctuations can be
expressed as a combination of generalized hypergeometric functions.
The cumulants of those distributions are, in general dimension $d$:
\begin{equation}
\label{eq:localcumulant}
\langle \epsilon^p \rangle_c (x)= \frac{n(x)\Delta x T^p(x)}{2^p} 
d(d+2) \dots(d+2p-2)\,\,,
\end{equation}
where the notation $\langle X \rangle_c$ denotes the cumulant of the
variable $X$.
\subsection{Comparison with simulations}

\begin{figure}[t]
\begin{minipage}[t]{.46\linewidth}
 \includegraphics[clip=true,width=1 \textwidth]{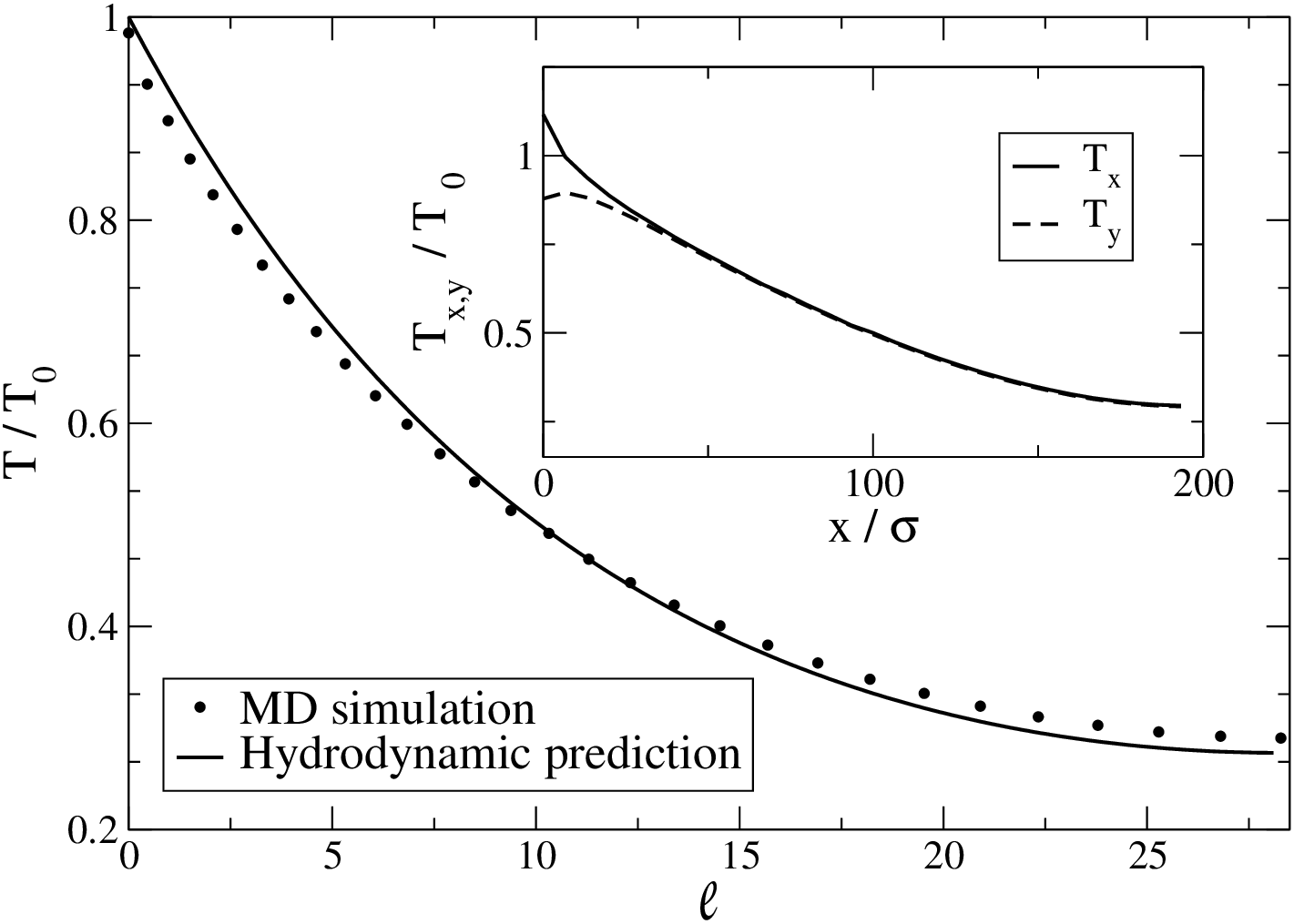}
\caption{\label{fig:profiles}  
  Temperature profile in the $\ell$ scale for a system of $N=1000$ particles
  in a box of height $H=200 \, \sigma$ and density $n \sigma^2=0.05$ with
  $\alpha=0.99$. The temperature is scaled with the temperature near the
  vibrating wall $T_0$. The inset show the temperature profile of the $x$ and
  $y$ component in the $x$ scale, showing that the temperature is effectively
  isotropic apart from a small region near the vibrating wall. }
\end{minipage}
\hfill
\begin{minipage}[t]{.46\linewidth}
  \includegraphics[clip=true,width=1
  \textwidth]{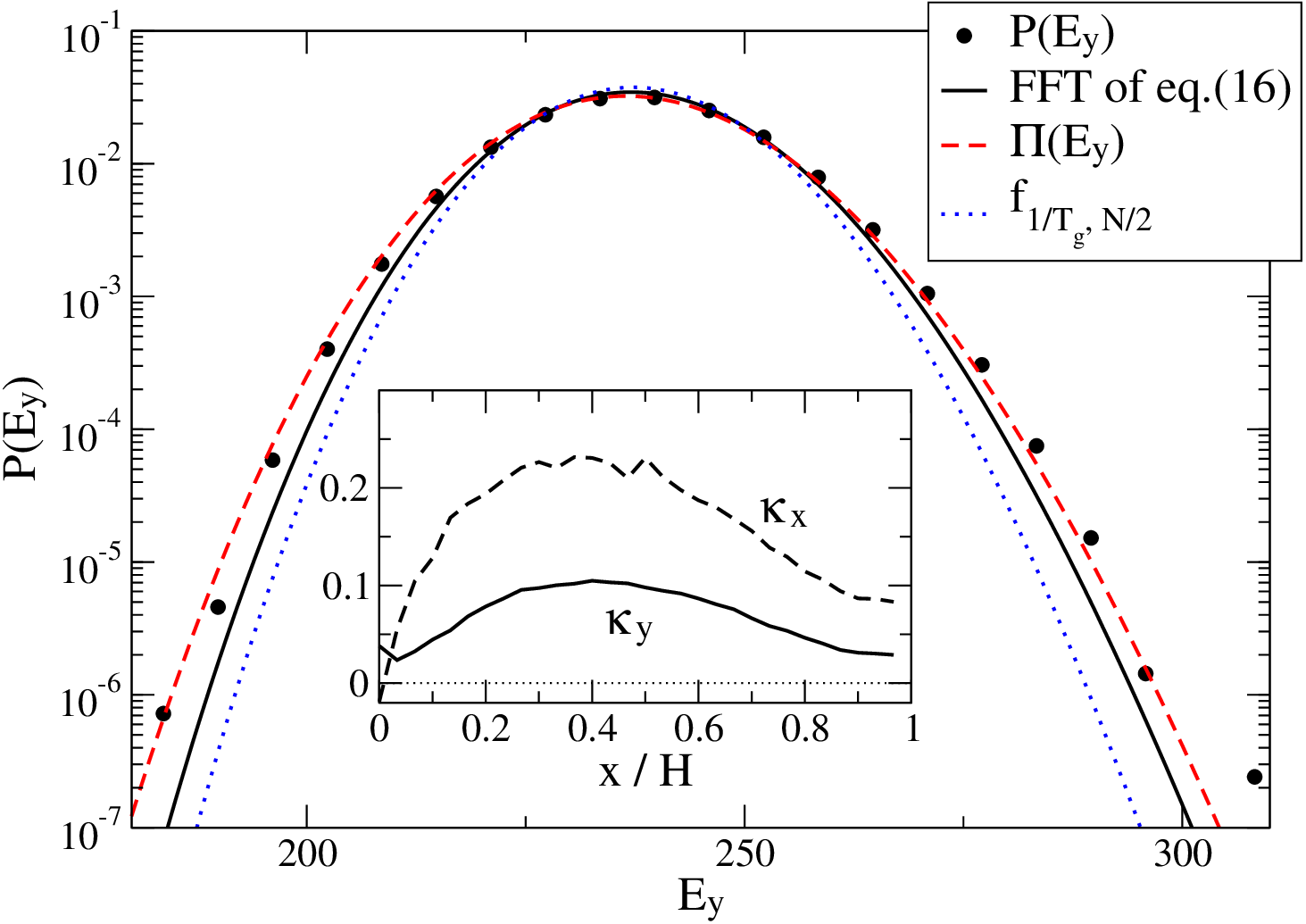}
  \caption{\label{figenpdf} Probability density function of the $y$ component 
    of the total energy $E_y=\sum_{i}^N v_{y_i}^2$ (dots). The parameters of
    the system are the same as in Fig. \ref{fig:profiles}.  The solid line is
    the numerical inverse Fourier transform of Eq.~(\ref{charfunen}), while
    the dashed line is a $\chi^2$ distribution with same mean and same
    variance of the numerical data, and the dotted line a 
  $\chi^2$ distribution with parameters corresponding to the granular
temperature and the true number of degrees of freedom. 
The energy is in units of the temperature
    of the wall $T_0$. The inset shows the kurtosis excess $\kappa_i=\langle
    v_i^4 \rangle/\langle v_i^2 \rangle^2 -3$ for the $x$ and $y$ component of
    the velocity pdf.}
\end{minipage}
\end{figure}


To obtain the analytic form of the energy pdf one should calculate the inverse
Fourier Transform of equation (\ref{charfunen}) using (\ref{tempprofile}) as
temperature profile. This does not seem possible analytically, but one can
resort to numerical procedures.
Auma\^itre {\it et al.}  \cite{aumaitre00a,aumaitre04} have shown by Molecular
dynamic simulations that the energy pdf is very well fitted by a $\chi^2$ law,
but with a number of degrees of freedom different from $N \,d$, and a
temperature different from the granular temperature $T_g$. They proposed to
define an effective number of degrees of freedom $N_f$ and an effective
temperature $T_E$, such that energy pdf should be of the form $\Pi
(E) = f_{{1 \over T_E}, {N_f \over 2}} (E)$, with:
\begin{equation}
N_f= 2 \, { \langle E \rangle_c^2  \over {\langle E^2 \rangle_c}}\,, 
\;\;\;\;\;\;\;\;\;\;\;\;\;\;\;\;\;\;\;\;
T_E= {\langle E^2 \rangle_c \over \langle E \rangle_c}\,\,.
\end{equation}
In Fig.~\ref{fig:profiles} a direct comparison between the measured
temperature profile in the $\ell$-scale and the theoretical prediction of
Eq.~(\ref{tempprofile}) shows the agreement between the Molecular Dynamics
(MD) simulation and the hydrodynamic prediction.  In Fig.~\ref{figenpdf}, the
Inverse Fourier Transform of (\ref{charfunen}) is compared with the pdf of the
$y$ component of the total energy obtained from MD simulations, showing a good
agreement. We restrict our analysis to the $y$ component because the velocity
pdf in this direction is closer to the Gaussian distribution (as quantified in
the inset by the excess kurtosis, see also~\cite{brey00c}), and is therefore
in better agreement with our hypotheses. In the same figure we have also
plotted the function $\Pi (E)$ previously defined, which has a similar shape
and yields a better fit of the numerical data. We emphasize however that our
prediction needs only one fitting parameter, $T_0$, which essentially fixes
the average of the distribution, while in order to have a closed form for $\Pi
(E)$ one needs to know both the mean and the variance of the distribution.
This underlines the importance of the inhomogeneous hydrodynamic profile in
determining the fluctuations of the global energy; on the other hand, the
correlations, which have been neglected in our approximate approach, seem to
play a lesser role.

Our approach also allows to investigate the behavior of the macroscopic
quantities $N_f$ and $T_E$ when the system size is changed.  Indeed, an
expression of the cumulants of the total kinetic energy can be obtained from
the characteristic function (\ref{charfunen}):
\begin{equation}
\label{energymom}
\langle E^p \rangle_c= {N d \over 2 \ell_m} \int_0^{\ell_m} T^p (\ell)
\textrm{d} \ell \,\,\,,
\end{equation}
so that
\begin{equation}
N_f= \frac{N d}{\ell_m} 
\frac{\left( \int_0^{\ell_m} T(\ell)d\ell \right)^2}
{\int_0^{\ell_m} T^2(\ell)d\ell }\,,\;\;\;\;\;
T_E= \frac{\int_0^{\ell_m} T^2(\ell)d\ell }
{\int_0^{\ell_m} T(\ell)d\ell } \  ,
\end{equation}
which shows that $T_E$ is not rigorously intensive but depends on the number
of layers of particles through $\ell_m$. At large enough $\ell_m$ however, the
integral in equation~\eqref{energymom} becomes size independent:
\begin{equation}
\int_0^{\ell_m} T^p(\ell) \textrm{d}\ell  \sim {T_0^p \over 2 p
  \sqrt{a(\alpha)}} \,\,,
\end{equation}
so that the effective temperature $T_E$ defined above becomes a constant
proportional to the temperature of the wall, while the parameter $N_f$ still
depends on the system size:
\begin{equation}
N_f \sim  {1 \over \sqrt{a(\alpha)}} \, {N d\over  \ell_m}\,,
\;\;\;\;\;\;\;\;\;\;\;\;\;\;\;\;\;\;\;\;
T_E\sim {T_0 \over 2}.
\label{asympt}
\end{equation}
We now investigate how the numerical results of~\cite{aumaitre04} fit with the
previous framework, for the various possibilities of system size variations at
fixed particle size $\sigma$ and constant density $\rho=N/(H L)$ (we consider
for simplicity a two-dimensional system, as in \cite{aumaitre04}):
\begin{itemize}
  
\item if only $L$ is increased, at constant $H$, $N/L$ is constant so that
  $\ell_m$ and thus $T_E$ do not change

\item if only $H$ is increased, at constant $L$, $\ell_m\propto N/L$ increases
  so that $T_E$ is not constant. The variation of $T_E$ with $H$ is however
  slow and moreover, as soon as $N$ is large enough, $T_E$ reaches its
  asymptotic value $T_0/2$.
  
\item if both $H$ and $L$ are increased at the same rate, $\ell_m$ increases
  slower so that once again $T_E$ is not rigorously constant but varies slowly
  towards $T_0/2$.
\end{itemize}

\begin{figure}[t]
\begin{minipage}[t]{.46\linewidth}
 \includegraphics[clip=true,width=1 \textwidth]{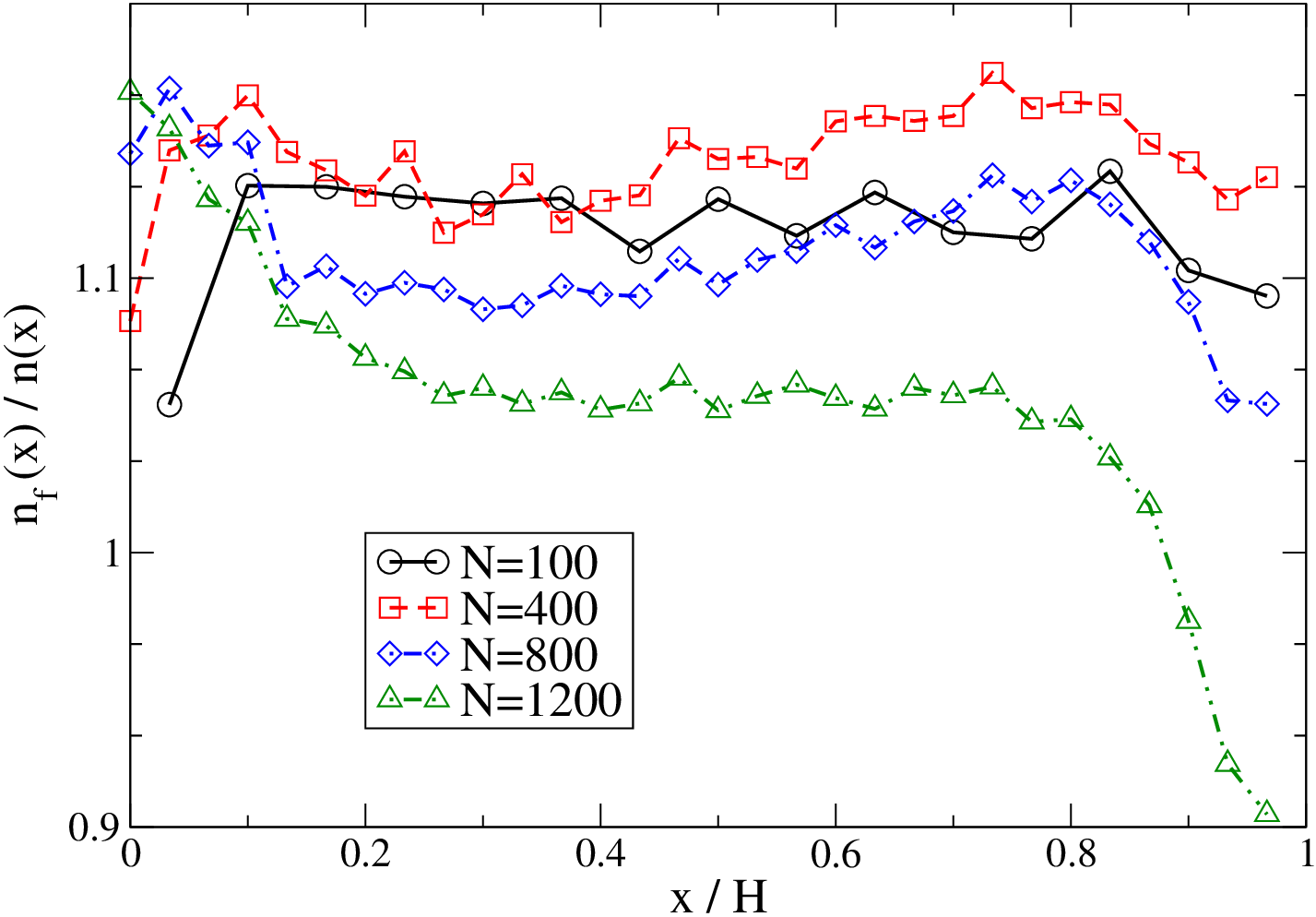}
\caption{\label{fig:effectivedensity} 
  Plot of the effective density $n_f(x)$ divided by the true density
  $n(x)$ in a square box at constant density $\rho \sigma^2=0.04$ for
  $\alpha=0.99$ and several system sizes. The corresponding effective
  numbers of degrees of freedom are, when the number of particles and
  the volume are increased at constant density: $N_f/(N d)=0.99$;
  $N_f/(N d)=0.77$; $N_f/(N d)=0.63$; $N_f/(N d)=0.47$.}
\end{minipage}
\hfill
\begin{minipage}[t]{.46\linewidth}
  \includegraphics[clip=true,width=1 \textwidth]{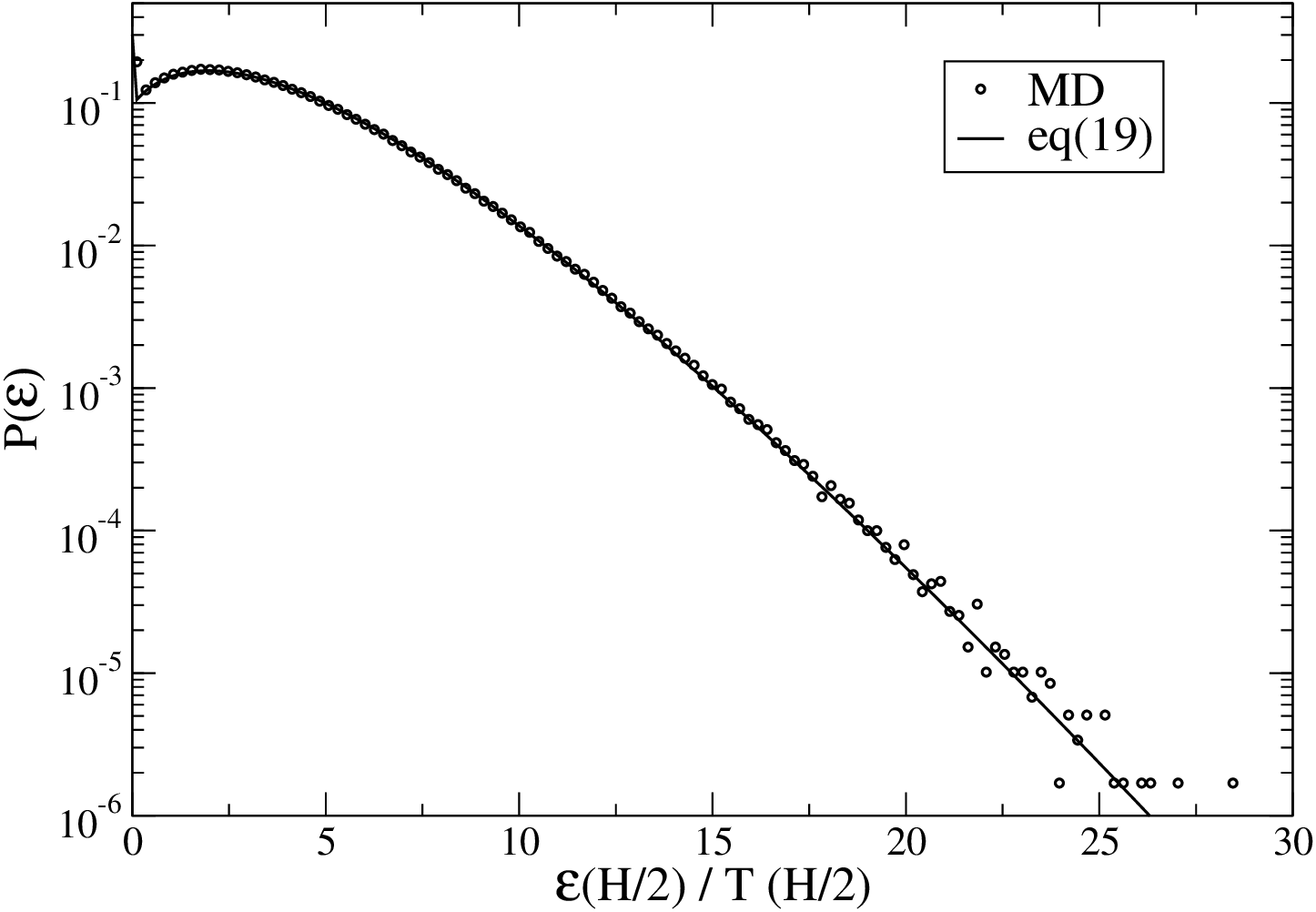}
  \caption{\label{fig:localenergypdf} Plot of the local energy pdf
    for a system of $N=100$ particles in a square cell at density
    $\rho \sigma^2 = 0.04$ and $\alpha=0.95$. The energy has been
    sampled in a strip of width $\Delta x = 1.66\,\sigma$ centered in
    $x=H/2$. The solid line is the analytical estimation from
    Eq.~(\ref{eq:pdfstripe2}).}
\end{minipage}
\end{figure}

The behavior of $N_f$ also depends on the maximum of the integrated density
$\ell_m$. At constant $\sigma$ and density $\rho=N/(H L)$, Eq.~(\ref{asympt})
shows that (i) for a square cell, since $\ell_m \propto \sqrt{N}$ , one
obtains $N_f \propto \sqrt{N}$, close to the $N^{0.4}$ behavior numerically
observed in \cite{aumaitre04}, and (ii) at constant $L$, if only the height
$H$ of the cell is increased, $\ell_m$ is proportional to $N$, and $N_f$ thus
becomes constant. All these features are in agreement with the numerical
observations in \cite{aumaitre04}. The above results show that the issues of
intensivity and extensivity can be understood through an approximate approach
which neglects correlations but focuses on the inhomogeneities as described by
the hydrodynamic profiles~\cite{brey00c}. The crucial quantity turns out to be
the number of layers of particles at rest. If this number is large enough, the
effective temperature $T_E$ becomes intensive. Moreover, this approach is
able to quantitatively describe the behavior of the fluctuations of the total
kinetic energy of a vibrated granular gas.  In some cases the energy pdf can
be approximated with a gamma distribution, which is the standard distribution
for the energy pdf in the canonical equilibrium.  Nevertheless there are
strong deviations from the equilibrium theory of fluctuations, since in this
case the two parameters of the gamma distribution (i.e. the temperature and
the number of degrees of freedom) are not the granular temperature nor the
total number of degrees of freedom. Another important remark is that
correlations, and in particular contributions from the two points distribution
function, do not play a primary role in explaining those deviations from the
equilibrium theory of fluctuations.  In order to characterize deviations which
do not arise from inhomogeneities, it can be useful to measure energy
fluctuations at a given height $x$ from the vibrating wall. From the
expression of the cumulants of the local energy pdf (\ref{eq:localcumulant})
it is possible to define an ``effective density'' (of particles) $n_f(x)=
\frac{d(d+2)}{d^2} \frac{\langle \epsilon \rangle^2}{\langle \epsilon^2
  \rangle_c}$. If the uncorrelated and Gaussian hypotheses are satisfied, then
$n_f(x)=n(x)$, and deviations from this identity should be due only to
correlations and non-Gaussianities. In Figs. \ref{fig:effectivedensity} and
\ref{fig:localenergypdf} results from MD simulations are shown at various
values of the inelasticity, density and number of particles. The agreement
between Eq.~(\ref{eq:pdfstripe2}) and the numerical data is very good, while
the deviations from $n_f(x)=n(x)$ shown in Fig.~\ref{fig:effectivedensity},
are evident but small compared to the deviations between $N_f$ and $N$.

Another way to discriminate the respective role of heterogeneities
and of correlations for the global system is to 
study granular gases maintained in an
homogeneous state by a suitable energy injection, as proposed below.

\section{The homogeneously driven case}


In this section we consider a granular gas kept in a stationary state by an
external homogeneous thermostat. Two different kinds of thermostat will be
used. We will first consider the so called ``Stochastic thermostat'', which is
a white noise acting independently on each particle
\cite{williams96b,noije98c,puglisi99,noije99,montanero00b,moon01,garzo02b}. In
a second part we will show some numerical results concerning the ``Gaussian
thermostat'', which consists of a negative friction force acting on each
particle \cite{montanero00b,garzo02b}.

\subsection{Stochastic thermostat}
\begin{figure}[t]
\begin{minipage}[t]{.46\linewidth}
 \includegraphics[clip=true,width=1 \textwidth]{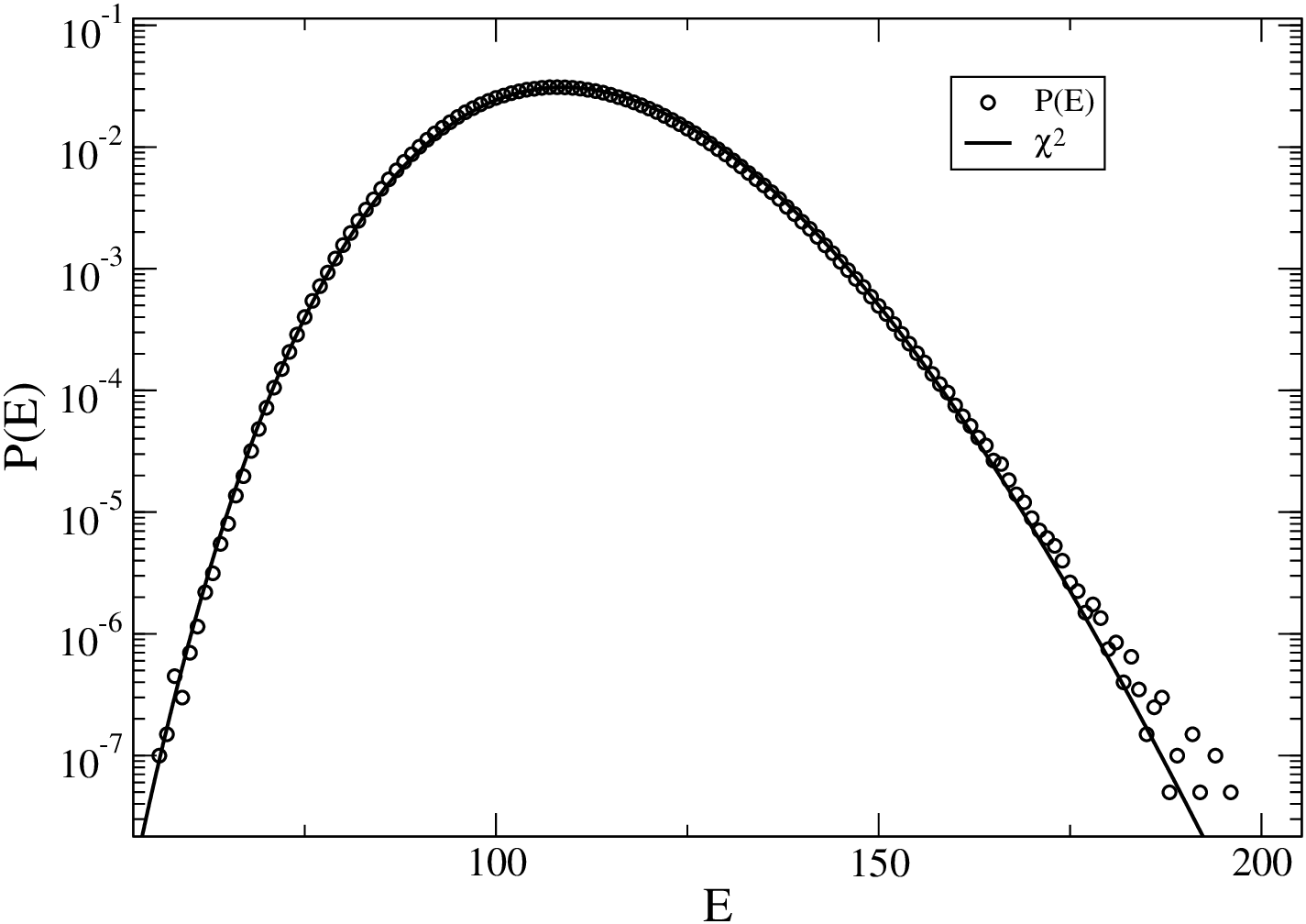}
\caption{\label{figenpdfdriv} Energy pdf (dots) from DSMC simulations with a 
    restitution coefficient $\alpha = 0.5$ and $N=100$ particles for a system
    driven with the stochastic thermostat. The solid line shows a gamma
    distribution with same mean and same variance.}
\end{minipage}
\hfill
\begin{minipage}[t]{.46\linewidth}
  \includegraphics[clip=true,width=1 \textwidth]{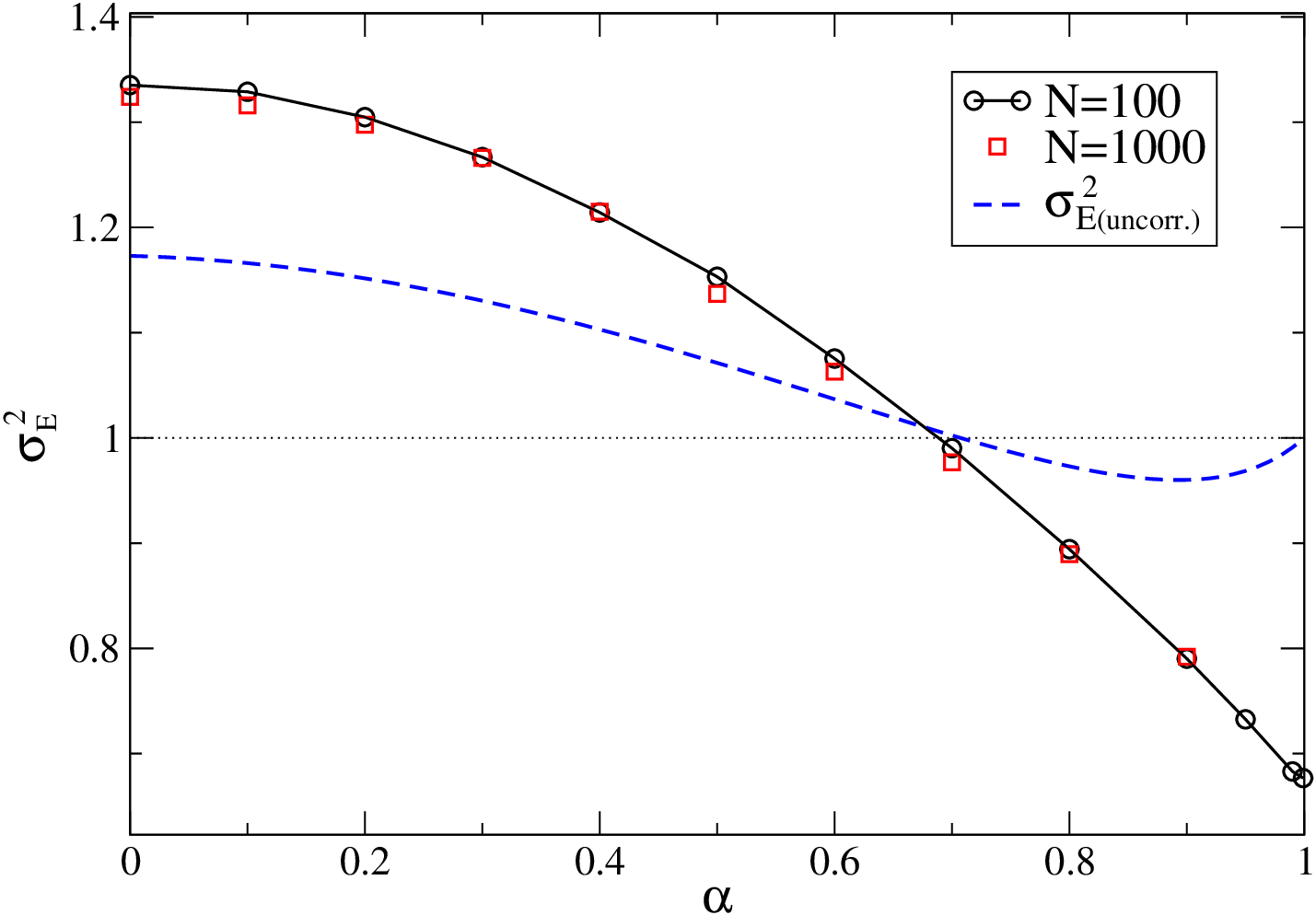}
\caption{\label{ealphadriven} Plot of $\sigma^2_E$ versus the restitution
  coefficient $\alpha$ for $N=100$ ($\bigcirc$) and $N=1000$
  (\textcolor{red}{$\square$}) particles driven by the stochastic thermostat. 
  The result of the calculation
  assuming uncorrelated velocities (\ref{sigmauncorr}) is plotted in dashed
  line.}
\end{minipage}
\end{figure}

We consider here a gas of $N$ inelastic smooth hard spheres driven by an
external random noise. The equation of motion of each particle $i$ is:
\begin{equation}
{\ddr \vvec_i \over \ddr t}={{\bf F}^{c}_i \over m} + \hat{{\boldsymbol
    \xi}}_i(t)\,,
\end{equation} 
where $\fvec^c_i$ is the force due to inelastic collisions [the latter being
ruled by Eqs. (\ref{coll})], and $\hat{{\boldsymbol \xi}}_i$ is the random
acceleration due to the stochastic force, which is assumed to be an
uncorrelated Gaussian white noise:
\begin{equation}
\langle {\hat \xi}_{i \alpha}(t) {\hat \xi}_{j \beta}(t') \rangle=\xi_0^2 
\delta_{ij} \delta_{\alpha \beta} \delta(t-t') \,\,,
\end{equation} 
where $i$ and $j$ denote the labels of the particles, while $\alpha$ and $
\beta$ refer to the Euclidean component of the noise. Our interest goes to the
stationary state of such system, which is homogeneous, and furthermore does
not develop spatial instabilities. A good description of such a system can
therefore be obtained through a homogeneous Boltzmann equation, which will
contain a Fokker-Planck diffusion term, taking into account the stochastic
force. This equation reads \cite{noije98c}:
\begin{equation}
\label{boltz}
\partial_t f({\bf v}_1, t)= J[f|f] + {\xi_0^2 \over 2} \left({\partial \over
    \partial \vvec_1} \right) f(\vvec_1, t)\,\,,
\end{equation} 
where $J[f|f]$ is the collision integral (cf. Eq.~\ref{collint}).  The
stationary solution of this equation is well approximated by the product of a
Gaussian with a Sonine Polynomial
\begin{equation}
\label{pdfsonine}
f(\cvec)={e^{-c^2} \over \pi^{d/2}}(1+ a_2 S_2(c^2))\,\,,
\end{equation}
where $\cvec = \vvec /v_0$ is a dimensionless velocity,
with $v_0=\sqrt{2 T}$, $S_2 (x)$ is the second Sonine polynomial
\begin{equation}
S_2(x)={1 \over 2} x^2 -{1 \over 2} (d+2)x +{1 \over 8} d(d+2) \,\,,
\end{equation}
and $a_2$ is a coefficient proportional to the kurtosis of the function
$f(\cvec)$:
\begin{equation}
a_2={4 \over d (d+2)} \left[ \langle c^4 \rangle - 
{1 \over 4} d (d+2) \right] = {4 \over 3} \left[\langle c_x^4 \rangle  
-3 \langle c_x^2 \rangle^2\right]  \,\,.
\end{equation}
An approximate expression of the coefficient $a_2$ for an arbitrary
restitution coefficient is \cite{noije98c}:
\begin{equation}
a_2 (\alpha)={16 (1- \alpha)(1-2 \alpha^2) \over 73 +56 d 
-24 \alpha d -105 \alpha +30 (1-\alpha)\alpha^2}\,\,.
\end{equation}

The energy dissipated in the inelastic collisions is compensated by the energy
injected by the thermostat so that the system reaches a stationary state, in
which the temperature fluctuates around its mean value
\begin{equation}
T_g=m \left(\frac{d \xi_0^2 \sqrt{\pi}}{(1-\alpha^2) \Omega_d n \sigma^{d-1}} 
\right)^{2/3}(1 + {\cal O} (a_2))\,\,,
\end{equation} 
where $\Omega_d$ is the $d$-dimensional solid angle, $m$ the mass of the
particle, and $n$ the density of the system. Here we are interested in the
fluctuations of the total energy measured by the quantity
\begin{equation}
\label{eq:sigmae}
\sigma_E^2=N {\langle E^2(t) \rangle - \langle E(t) \rangle^2 \over \langle
  E(t) \rangle^2}\,\,.  
\end{equation}
Note that $\sigma_E^2\equiv 2N/N_f$, and that $T_g$ is intensive while
the total energy is extensive. Brey {\it et al.} have computed, by
means of kinetic equations, an analytical expression for $\sigma_E^2$
in the homogeneous cooling state, which is equivalent to the so-called
Gaussian deterministic thermostat~\cite{brey04a}. This quantity has
also been computed in a one dimensional granular gas with a similar
thermostat \cite{cecconi2004}. One of the main differences of the
stochastic thermostat with a deterministic one is found in the elastic
limit.  On the one hand, for the cooling state, when the restitution
coefficient tends to $1$, the conservation of energy imposes that the
energy pdf is a Dirac delta function, and the quantity $\sigma_E$ goes
to $0$. On the other hand, with the stochastic thermostat, if the
elastic limit is taken keeping the temperature constant, the strength
of the white noise will tend to zero, but it will still play a role in
the velocity correlation function (we note that it is not necessary to
keep the temperature constant while performing the limit, since this
quantity only sets an irrelevant time scale).

An efficient way to have numerical solutions of Boltzmann-like equations is
the so called Direct Simulation Monte Carlo (DSMC). This algorithm simulates a
Markov chain whose associated master equation is expected to converge to the
Boltzmann equation.  We performed this kind of simulation to measure the
energy pdf. A plot of this quantity is shown in Fig.~\ref{figenpdfdriv}. We
observe again that it is close to a $\chi^2$-distribution with same mean and
same variance.  Nevertheless the number of degrees of freedom of this
$\chi^2$-distribution is lower than the true number of degrees of freedom
(i.e. $(N-1) \times d $).  This effect may arise from two separated causes:
the non-gaussianity of the velocity pdf, and the presence of correlations
between the velocities. This feature also suggests that a calculation of the
energy pdf with the hypothesis of uncorrelated velocities (but non-Gaussian)
could explain at least a part of this non-trivial effect. The calculation of
the pdf of the sum of the square of $n$ variables distributed following
(\ref{pdfsonine}) is straightforward. The characteristic function of the
energy pdf is:
\begin{equation}
{\widetilde P}_N(k)= {1 \over (1-\im k T)^{Nd \over 2}} 
\left(1+{d (d+2) \over 8} a_2 
\left({1 \over (1- \im k T)^2}- {2 \over (1-\im k T)} +1 \right) 
\right)^{N}
\end{equation}
where $N$ is the number of particles of the system.  This results yields
\begin{equation}
\langle E \rangle = {d \over 2} N T\,\,,
\;\;\;\;
\langle E^2 \rangle - \langle E \rangle^2={d \over 2} N T^2 \left(1+{d+2 \over
    2} a_2 \right) \,\,,
\end{equation}
so that the explicit expression for the energy fluctuations reads:
\begin{equation}
\label{sigmauncorr}
\sigma^2_{E_{(uncorr.)}}={2 \over d} \left(1+{d+2 \over 2} a_2 \right) \,\,.
\end{equation} 
As expected, the temperature does not appear in the dimensionless $\sigma_E^2$
above, which depends on the only available dimensionless parameters $\alpha$
and $d$. This result is compared in Fig.~\ref{ealphadriven} with the result of
DSMC simulations, carried out for several values of the restitution
coefficient $\alpha$ and for two different values for the number of particles
$N$. The disagreement between the uncorrelated calculation and the simulations
is a clear sign of the correlations induced by the inelasticity of the system
(the tail of the velocity pdf is also non-Gaussian \cite{noije98c}, but this
feature has negligible consequences for the quantities of interest here).  One
can note that the fluctuations increase when the restitution coefficient
decreases. One can also see that there is a value of the restitution
coefficient $\alpha$ around $1/\sqrt{2}$, that is when the approximate
expression of $a_2$ vanishes, for which $\sigma_E^2$ is exactly $1 \equiv
2/d$, as for a gas in the canonical equilibrium.

\begin{figure}[t]
  \begin{minipage}[t]{.46\linewidth}
  \includegraphics[clip=true,width=1. \textwidth]{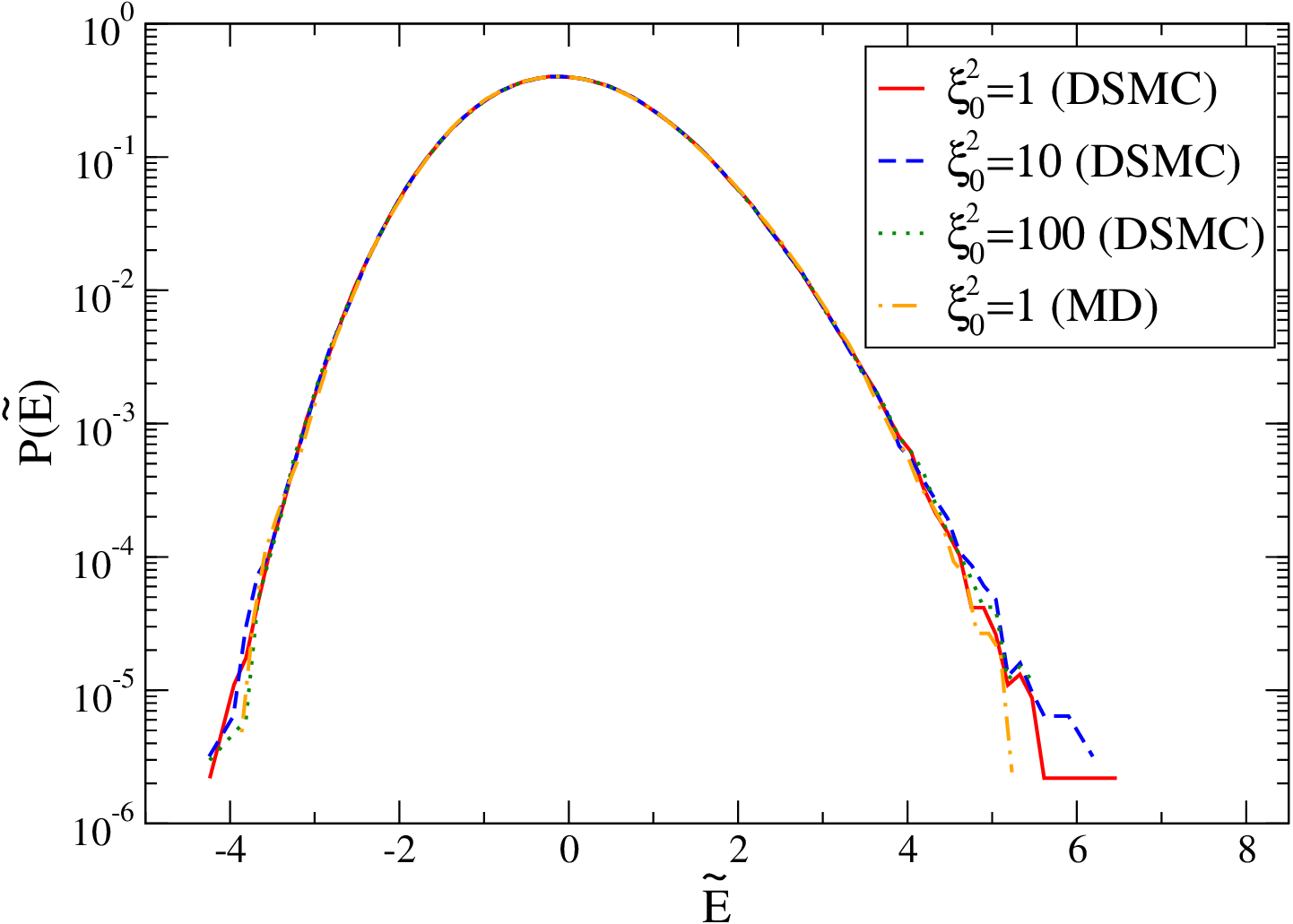}
  \caption{\label{allrescaled} Plot of the pdf of ${\widetilde E}$ for a 
    restitution coefficient $\alpha=0.5$, for $N=100$ particles, and
    for several values of the noise's strength $\xi_0^2$. The
    simulations have been carried on with both Monte-Carlo (DSMC) and
    Molecular Dynamics (MD) algorithms, with $n\sigma^2=0.04$.}
\end{minipage}
\hfill
\begin{minipage}[t]{.46\linewidth}
  \includegraphics[clip=true,width=1 \textwidth]{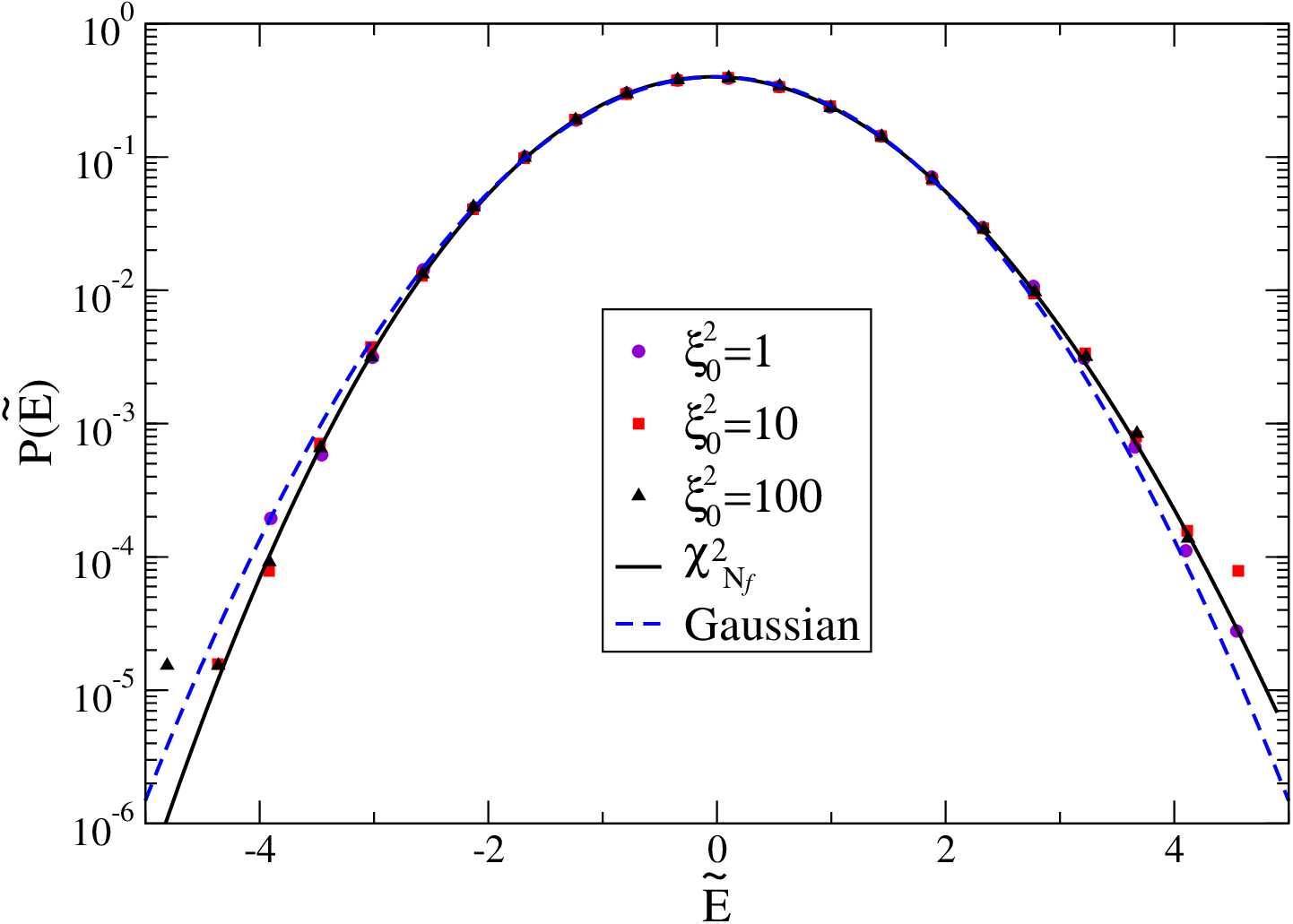}
\caption{\label{allrescalebis} Plot of the pdf of ${\widetilde E}$ for a 
  restitution coefficient $\alpha=0.5$, for $N=1000$ particles, and for
  several values of the noise's strength $\xi_0^2$. The solid line is a
  rescaled $\chi^2$-distribution with $N_f=2 N/\sigma_E^2 \simeq 1759$ degrees
  of freedom, and it is very close to the Gaussian distribution, plotted here
  in dashed line.}
\end{minipage}
\end{figure}

We now turn to the dependence of $\sigma^2_E$ on the strength of the white
noise $\Gamma$. It is useful, for this purpose, to introduce a rescaled,
dimensionless energy
\begin{equation}
{\widetilde E}={E - \langle E \rangle \over \sqrt{\langle E^2 \rangle -
    \langle E \rangle^2}} \,\,.
\end{equation} 
We have plotted in Fig.~\ref{allrescaled} this rescaled energy pdf for a
system of $N=100$ particles with a restitution coefficient $\alpha=0.5$ and
for several values of the strength of the white noise $\Gamma$. As expected,
all the pdfs collapse into a unique distribution: the role of the noise's
strength is only to set the temperature (or mean kinetic energy) scale.
Besides, relative energy fluctuations depend only on $\alpha$ and $N$.
Moreover, since $\sigma_E^2$ does not depend on the number of particles $N$
(for $N$ large enough), $\sqrt{\langle E^2 \rangle - \langle E \rangle^2}$
grows as $\sqrt{N}$, so that the central limit theorem applies, and hence
$P({\tilde E})$ is a Gaussian in the thermodynamic ($N \to \infty$) limit.
Figure \ref{allrescalebis} indeed shows how the rescaled energy
pdf for $N=1000$ particles is very close to a Gaussian, even if the 
$\chi^2$-distribution still gives a better fit.


\subsection{Gaussian thermostat}

In this section, we will focus on the fluctuations in a granular gas driven by
a Gaussian thermostat \footnote{This is the traditional nomenclature to denote
  such a thermostat for granular gases. Note, however, that the thermostating
  force acting on the particles is not obtained in such a way to satisfy
  Gauss' principle of least constraint. The total kinetic energy is not
  constant, but fluctuates in time. Henceforth we will refer to the Gaussian
  thermostat without any connection to Gauss' principle of least constraint,
  nor to the thermostats described in \cite{evansmorriss}.}, which consist in
adding a viscous force, with a negative friction coefficient, in the equation
of motion of each particle. Thus the velocity of the $i$-th particle will
verify the following equation:
\begin{equation}
\label{eqmotgauss}
\frac{\dd \bv_i}{\dd t}= \frac{\Fvec^c_i}{m} + \gamma \bv_i \,\,. 
\end{equation}
In the above equation $\bF^c$ is the force due to collisions, and $\gamma$ is
a positive constant. The Boltzmann equation corresponding to
Eq.~(\ref{eqmotgauss}) is:
\begin{equation}
\label{boltzgauss}
\partial_t f(\bv_1,t)=J[f|f]- \gamma \frac{\partial}{\partial \bv_1} \cdot 
[\bv_1 f(\bv_1, t)]\,\,\,.
\end{equation}
In this case also, the relevant solution is well approximated by a Gaussian
multiplied with a Sonine polynomial [cf.  Eq.~(\ref{pdfsonine})]. The
expression of the coefficient $a_2$ has been widely studied
\cite{noije98c,montanero00b,coppex03}, and the one which is in better
agreement with numerical simulations is the following \cite{montanero00b}:
\begin{equation}
\label{a2gauss}
a_2 (\alpha)=\frac{16 (1-\alpha^2)(1-2 \alpha^2)}
{25+24 d -\alpha(57-8 d)-2 (1-\alpha) \alpha^2}\,\,.
\end{equation}
The parameter $\gamma$ determines the strength of the homogeneous force acting
on the particles. Its primary role is to set the temperature scale of the
stationary state. Projecting Eq.~(\ref{boltzgauss}) on the second velocity
moment is it possible to get an approximate expression of the stationary
temperature:
\begin{equation}
\label{temperaturegauss}
T 
\simeq m \left(\frac{2 \sqrt{\pi} \gamma d}{(1- \alpha^2) \Omega_d n 
\sigma^{d-1}} \right)^2(1 + {\cal O} (a_2))\,\,.
\end{equation}
This result has been obtained assuming Gaussian velocity pdf, since the
corrections coming from the coefficient $a_2$ are small.

\begin{figure}
  \begin{minipage}[t]{.46\linewidth}
    \includegraphics[clip=true,width=1 \textwidth]{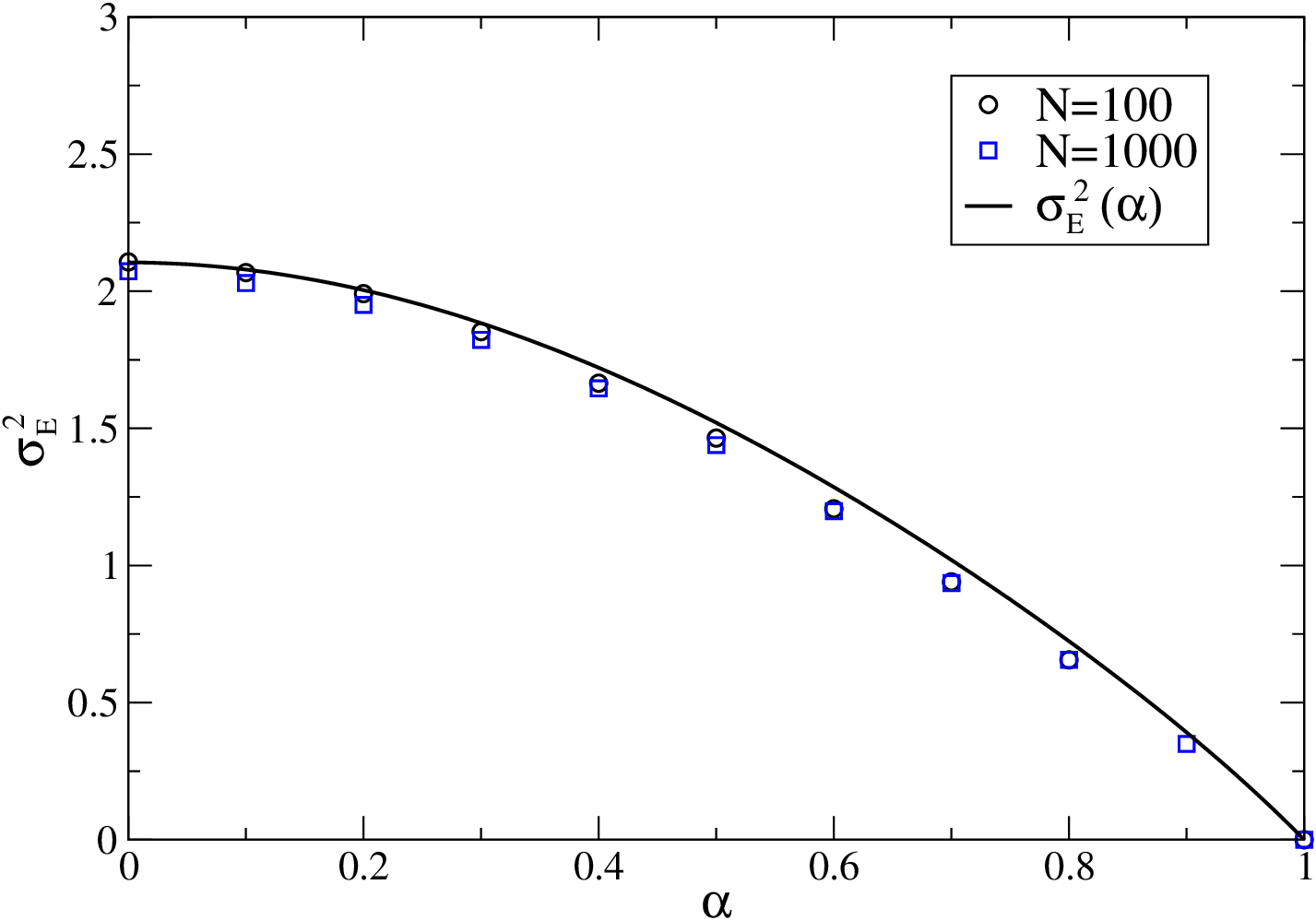}
    \caption{\label{ealphacooling} Plot of $\sigma^2_E$ versus the restitution
      coefficient $\alpha$ for $N=100$ ($\bigcirc$) and $N=1000$
      (\textcolor{blue}{$\square$}) particles, for the Gaussian thermostat.
      The solid line shows the theoretical predictions of \cite{brey04a},
      given by Eq.~(\ref{eq:sigmaegauss}).}
  \end{minipage}
  \hfill
  \begin{minipage}[t]{.46\linewidth}
    \includegraphics[clip=true,width=1
    \textwidth]{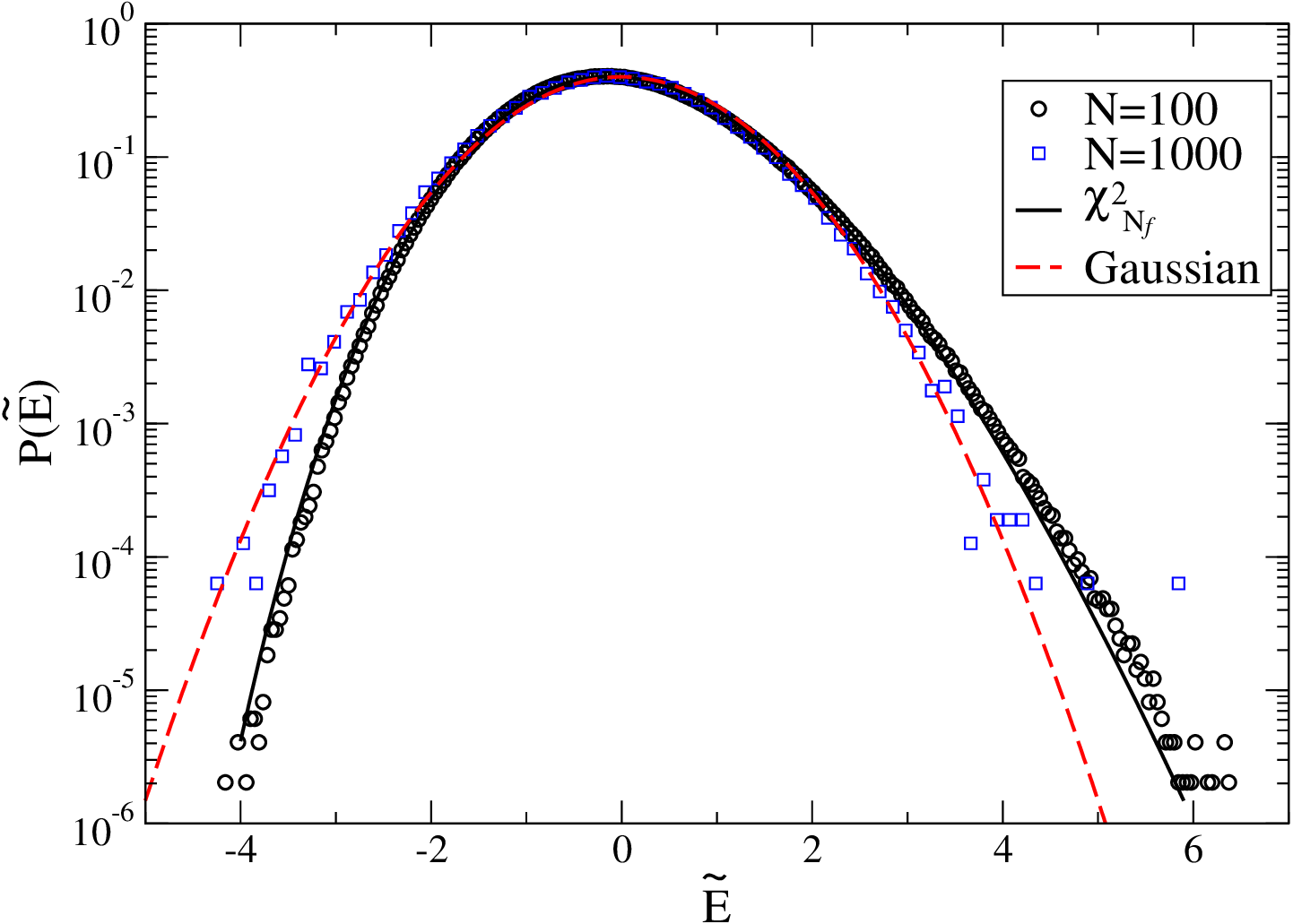}
    \caption{\label{figenpdfcool} pdf of the rescaled energy 
      $\widetilde E$ when the restitution coefficient is $\alpha = 0.5$.  The
      system is driven with the Gaussian thermostat with 100 ($\bigcirc$) and
      1000 (\textcolor{blue}{$\square$}) particles.  The solid line shows a
      rescaled gamma distribution with zero-mean, a variance equal to 1, and
      with a number of degrees of freedom $N_f$ given by $2 N /\sigma_E^2$.
      The dashed line is the Gaussian with zero-mean and unit variance.}
  \end{minipage}
  \hfill
\end{figure}

Equation~(\ref{boltzgauss}) is formally equivalent to the Boltzmann equation
for the homogeneous cooling state (HCS) of a granular gas, for the particular
solution in which all the time dependence of the velocity pdf appears trough
the time dependent temperature \cite{goldshtein95}. This mapping of the
cooling state onto a steady state can be obtained in an equivalent fashion
applying a logarithmic time-scale transformation to the time-dependent
Boltzmann equation of the HCS~\cite{lutsko01,brey04a}. Furthermore, the above
scaling solution can be extended to the $n$-particle distribution function of
the system.  The resulting hierarchy equations of the HCS are then exactly the
same one would obtain for a homogeneous granular gas thermostated by a
Gaussian thermostat. Thus these two models not only have the same velocity
pdf, but also the same $n$-particles distribution, and hence they will have
the same relative fluctuations and correlation functions.  Henceforth we will
no more make any distinction between the HCS and the Gaussian thermostat. We
will therefore compare results obtained for the HCS with numerical simulations
of a gas driven in a stationary state with a Gaussian thermostat.  We also
have to note that this kind of homogeneous state is unstable under small
density fluctuations, and naturally develops cluster and shear instabilities.
In the following we will therefore focus only on homogeneous states. For the
HCS Brey {et al.} \cite{brey04a} have obtained an analytical expression for
the quantity $\sigma_E^2$ defined in Eq.~(\ref{eq:sigmae}). Its expression
is:
\begin{equation}
\label{eq:sigmaegauss}
\sigma_E^2=\frac{d (d+1)}{2}+ \frac{d(d+2)}{4} a_2(\alpha) +d^2 b(\alpha)\,\,,
\end{equation}
where\footnote{Note that the following expression is not exactly the same
  given in \cite{brey04a}, since here a different expression of the
  coefficient $a_2$ (which fits better with numerical simulations) is used.}
\begin{equation}
  \label{eq:b}
  b(\alpha)= \frac{4\,\left( 1 - 2\,\alpha^2 \right) \,
    \left( 1 - 3\,\alpha + \alpha^2 \, \left(\alpha - 1 \right) \right)
    - \left( 1 - \alpha \right) \,
    \left( 3 - 2\,\alpha\,\left( 8 + 5\,\alpha \right)  \right) \,d
    - 2\,\left( 3 + \alpha \right) \,d^2}{2\,d\,
    \left( 7 + 6 \, d - \alpha\,\left( 15 + 
        2\,\alpha \left( 1 - \alpha \right) - 2\,d \right) \right) }\,\,.
\end{equation}
Note that $b(1)=-(1+d)/(2d)$ so that $\sigma_E^2 \to 0$ as $\alpha \to 1$.
The results from DSMC simulations and the theoretical predictions by Brey {\it
  et al.} are shown in Fig.~\ref{ealphacooling}.  The simulations have been
carried out for two different sizes of the system ($N=100$ and $N=1000$), and
for several values of $\alpha$, and in each case the numerical results are in
good agreement with the predictions of the relation (\ref{eq:sigmaegauss}).
The energy pdf behaves again as a $\chi^2$ distribution. When the system is
not too large ($N \sim 100$) the shape of the energy pdf is very well fitted
with a $\chi^2$ distribution with a mean value obtained from the temperature
expression (\ref{temperaturegauss}) and a variance obtained from
(\ref{eq:sigmaegauss}). Furthermore, when the number of particles is
increased, the $\chi^2$-distribution becomes closer and closer to the Gaussian
distribution. All these features are shown in Fig.~\ref{figenpdfcool}. Ref.
\cite{brey04a} reported a good agreement between Molecular Dynamics results
and Eq.~(\ref{eq:sigmaegauss}). The fact that the relevant correlations are
already captured by the DSMC scheme is a remarkable feature that will be
commented further in the conclusion.

\section{Conclusion}

In this paper, we have investigated the fluctuations of the total kinetic
energy of granular gases maintained in a non-equilibrium stationary state
through various kinds of energy injection mechanisms.  For heating through a
boundary, the system remains inhomogeneous, and the characteristic function of
the energy pdf can be analytically computed as a functional of the temperature
profile of the system, under the assumption that the particle velocities are
uncorrelated. Despite this approximation, the use of the hydrodynamic profiles
obtained in \cite{brey00c} allows to recover results in agreement with the
numerical simulations carried out in Ref. \cite{aumaitre04}.  This result
seems to hold even if the hydrodynamic approximation is not valid in the whole
volume of the system, in particular near the vibrating wall. Moreover, we find
that the effective temperature defined in \cite{aumaitre04}, which is
measurable, is intensive only if the number of layers of particles is large
enough, and can be related to a quantity which plays the role of the
``temperature of the vibrating wall'', which is microscopically not well
defined, and {\it a priori} not measurable. An effective number of degrees of
freedom can be also defined, and its dependence on the true number of degrees
of freedom has been obtained. Once again, the obtained behavior is in
agreement with the numerical simulations described in \cite{aumaitre04}, and
leads to the conclusion that the non-extensive behavior of the total energy in
such a system is essentially due to the density and temperature
inhomogeneities and can be understood through a hydrodynamic approach.

In homogeneously driven granular gases the total energy displays a $\chi^2$
pdf, with a number of degrees of freedom proportional to the number of
particles. The total energy is hence extensive, and moreover, in the $N \to
\infty$ limit the $\chi^2$ distribution tends to the Gaussian. We measured, by
means of DSMC simulations, the rescaled variance $\sigma_E^2$ of the energy
pdf, which depends only on the restitution coefficient. These non-trivial
fluctuations are essentially due to the correlations induced by the
inelasticity of the particles. We can distinguish between two different
contributions to these correlations. First, the non-Gaussianity of the
velocity pdf, which simply tells that the Euclidean components of the velocity
of each particle are correlated one to each other. Second, a contribution from
the two particles velocity pdf, which does not factorize exactly as a product
of two one-particle distributions. The analytical computation of the rescaled
variance $\sigma_E^2$ under the assumption of uncorrelated non-Gaussian
individual velocity pdfs does not reproduce the DSMC results, showing the
relevance of this second contribution.  It must be pointed out, however, that
these correlations do not invalidate the Boltzmann equation. As already noted
in \cite{ernst81,brey04a}, the two points correlation function $g_2(\bv_1,
\bv_2)$, which is defined by:
\begin{equation}
 g_2(\bv_1, \bv_2)=f^{(2)}(\bv_1, \bv_2)-f(\bv_1)f(\bv_2)\,\,,
\end{equation}
where $f^{(2)} $ is the two points distribution, is of higher order in the
density expansion (roughly speaking ${\cal O} \left(g(\bv_1, \bv_2) \right)
\sim {\cal O} \left(f(\bv_1)f(\bv_2) \right)/N$). This is confirmed by the
numerical observation, since the energy pdf tends to a Gaussian when the
number of particles increases.

When the gas is heated with the Gaussian thermostat, the numerical results are
in good agreement with the analytical result in \cite{brey04a}. This means in
particular that the DSMC algorithm is able to capture effects arising from the
two-particle velocity pdf, which is remarkable. Moreover the energy pdf
obtained with DSMC coincides with its counterpart found in MD simulations in
the dilute limit, provided the system remains homogeneous (cf. Fig.
\ref{allrescaled}). This feature is consistent with the fact that the DSMC
algorithm exactly simulates the homogeneous pseudo-Liouville equation
governing the $N$-particle dynamics of the gas.

For the gas thermostated by a random noise, numerical simulations show that
the quantity $\sigma_E^2$ behaves in a fashion similar to what happens for the
Gaussian thermostat. Nevertheless in the elastic limit ($\alpha \to 1$) it
tends to a non trivial value ($\sim d/\sqrt{2}$), which is still not
understood. An analytical calculation of $\sigma_E^2$ is in principle
possible, exploiting the methods described in \cite{brey04a}. Nevertheless, in
order to carry on with such methods, one needs the expression of the
eigenfunctions of the linearized Boltzmann operator. While for the Gaussian
thermostat these eigenfunctions can be computed, in the way as for the
elastic hard sphere gas, this task seems more demanding in the case of a
stochastically driven granular gas.

\begin{acknowledgement}
  The authors thank S. Auma\^itre, S. Fauve and J. Farago 
 for stimulating discussions.
\end{acknowledgement}

\bibliographystyle{epj}
\bibliography{flupaper}

\begin{thebibliography}{35}

\bibitem{racz03}
Z.A. R\'acz, Proc. SPIE Int. Soc. Opt. Eng. \textbf{5112}, 248 (2003)

\bibitem{bramwell98}
S.T. Bramwell, P.C.W. Holdsworth, J.F. Pinton, Nature \textbf{396}, 552 (1998)

\bibitem{brey05}
J.~Brey, M.~de~Soria, P.~Maynar, M.~Ruiz-Montero, Phys. Rev. Lett. \textbf{94},
  098001 (2005)

\bibitem{barrat05}
A.~Barrat, E.~Trizac, M.H. Ernst, J. Phys. Condens. Matter \textbf{17}, S2429
  (2005)

\bibitem{poschel01}
T.~P\"oschel, S.~Luding, eds., \emph{Granular Gases} (Springer, Berlin, 2001),
  {L}ecture Notes in Physics 564

\bibitem{poschel03}
T.~P\"oschel, N.~Brilliantov, eds., \emph{Granular Gas Dynamics} (Springer,
  Berlin, 2003), {L}ecture Notes in Physics 624

\bibitem{losert99b}
W.~Losert, D.G.W. Cooper, J.~Delour, A.~Kudrolli, J.P. Gollub, Chaos
  \textbf{9}(3), 682 (1999), cond-mat/9901203

\bibitem{rouyer00}
F.~Rouyer, N.~Menon, Phys. Rev. Lett. \textbf{85}(17), 3676 (2000)

\bibitem{kudrolli00}
A.~Kudrolli, J.~Henry, Phys. Rev. E \textbf{62}(2), R1489 (2000),
  cond-mat/0001233

\bibitem{wildman02}
R.D. Wildman, D.J. Parkar, Phys. Rev. Lett. \textbf{88}(19), 064301 (2002)

\bibitem{brey00c}
J.J. Brey, M.J. Ruiz-Montero, F.~Moreno, Phys. Rev. E \textbf{62}, 5339 (2000)

\bibitem{bertin05}
E.~Bertin, Phys. Rev. Lett. \textbf{95}, 170601 (2005)

\bibitem{ernst81}
M.H. Ernst, E.G.D. Cohen, J. Stat. Phys \textbf{25}(1), 153 (1981)

\bibitem{brey04a}
J.~Brey, M.~de~Soria, P.~Maynar, M.~Ruiz-Montero, Phys. Rev. E
  \textbf{70}(011302) (2004)

\bibitem{grossman97}
E.L. Grossman, T.~Zhou, E.~Ben-Naim, Phys. Rev. E \textbf{55}, 4200 (1997)

\bibitem{mcnamara97}
S.~McNamara, J.L. Barrat, Phys. Rev. E \textbf{55}, 7767 (1997)

\bibitem{mcnamara98b}
S.~McNamara, S.~Luding, Phys. Rev. E \textbf{58}, 813 (1998)

\bibitem{kumaran98}
V.~Kumaran, Phys. Rev. E \textbf{57}(5), 5660 (1998)

\bibitem{barrat02b}
A.~Barrat, E.~Trizac, Phys. Rev. E \textbf{66}(5), 051303 (2002)

\bibitem{aumaitre04}
S.~Auma{\^i}tre, J.~Farago, S.~Fauve, S.~McNamara, Eur. Phys. J. B \textbf{42},
  255 (2004)

\bibitem{feller}
W.~Feller, \emph{Probability Theory and its Applications} (John Wiley \& Sons,
  New York, 1971)

\bibitem{abramowitz}
M.~Abramowitz, I.A. Stegun, \emph{Handbook of mathematical functions with
  formulas, graphs and mathematical tables} (Dover, New York, 1972)

\bibitem{aumaitre00a}
S.~Auma{\^i}tre, S.~Fauve, S.~McNamara, P.~Poggi, Eur. Phys. J. B \textbf{19},
  449 (2000)

\bibitem{williams96b}
D.R.M. Williams, F.C. MacKintosh, Phys. Rev. E \textbf{54}(1), R9 (1996)

\bibitem{noije98c}
T.P.C. van Noije, M.H. Ernst, Granular Matter \textbf{1}(2), 57 (1998)

\bibitem{puglisi99}
A.~Puglisi, V.~Loreto, U.M.B. Marconi, A.~Vulpiani, Phys. Rev. E \textbf{59},
  5582 (1999)

\bibitem{noije99}
T.P.C. van Noije, M.H. Ernst, E.~Trizac, I.~Pagonabarraga, Phys. Rev. E
  \textbf{59}, 4326 (1999)

\bibitem{montanero00b}
J.M. Montanero, A.~Santos, Granular Matter \textbf{2}(2), 53 (2000),
  cond-mat/0002323

\bibitem{moon01}
S.J. Moon, M.D. Shattuck, J.B. Swift, Phys. Rev. E \textbf{64}, 031303 (2001),
  cond-mat/0105322

\bibitem{garzo02b}
V.~Garzo, J.M. Montanero, Physica A \textbf{313}(3-4), 336 (2002)

\bibitem{cecconi2004}
F.~Cecconi, F.~Diotallevi, U.M.B. Marconi, A.~Puglisi, J. Chem. Phys.
  \textbf{120}, 35 (2004)

\bibitem{evansmorriss}
D.J. Evans, G.P. Morriss, \emph{Statistical Mechanics of Nonequilibrium
  Liquids} (Academic Press, London, 1990)

\bibitem{coppex03}
F.~Coppex, M.~Droz, J.~Piasecki, E.~Trizac, Physica A \textbf{329}, 114 (2003)

\bibitem{goldshtein95}
A.~Goldshtein, M.~Shapiro, J. Fluid Mech. \textbf{282}, 75 (1995)

\bibitem{lutsko01}
J.F. Lutsko, Phys. Rev. E \textbf{63}, 061211 (2001)

\end{thebibliography}

\end{document}